\documentclass[aps,pra,reprint,superscriptaddress]{revtex4-1}

\usepackage{amsmath}
\usepackage{graphicx}
\usepackage{times}
\usepackage{amssymb}
\usepackage{hyperref}
\usepackage{textcomp}
\usepackage{xcolor}
\usepackage[T1]{fontenc}
\usepackage{ulem}

\begin{document}

\title{Estimating the single-photon projection of low-intensity light sources}

\author{Jorge Rolando Chavez-Mackay}
\affiliation{Tecnologico de Monterrey, Escuela de Ingenier\'ia y Ciencias, Ave. Eugenio Garza Sada 2501, Monterrey, N.L. 64849, M\'exico.}
\author{Peter Gr\"unwald}
\affiliation{Aarhus Universitet, Institut for Fysik og Astronomi, Ny Munkegade 120, 8000 Aarhus C, Denmark.}
\author{Blas Manuel Rodr\'iguez-Lara}
\affiliation{Tecnologico de Monterrey, Escuela de Ingenier\'ia y Ciencias, Ave. Eugenio Garza Sada 2501, Monterrey, N.L. 64849, M\'exico.}
\affiliation{Instituto Nacional de Astrof\'{\i}sica, \'Optica y Electr\'onica, Calle Luis Enrique Erro No. 1, Sta. Ma. Tonantzintla, Pue. CP 72840, M\'exico.}

\begin{abstract}
	Estimating the quality of a single-photon source is crucial for its use in quantum technologies. 
	The standard test for semiconductor sources is a value of the second-order correlation function of the emitted field below $1/2$ at zero time-delay. 
	This criterion alone provides no information regarding the amplitude of the single-photon contribution for general quantum states.
	Addressing this question requires the knowledge of additional observables. 
	We derive an effective second-order correlation function, strongly connected to the Mandel-$Q$ parameter and given in terms of both the second-order correlation and the average photon number, that provides a lower bound on the single-to-multi-photon projection ratio.
	Using both observables individually allows for lower and upper bounds for the single-photon projection.
	Comparing the tightness of our bounds with those in the literature, we find that relative bounds may be better described using the average photon number, while absolute bounds for low excitation states are tighter using the vacuum projection. 
	Our results show that estimating the quality of a single-photon source based on additional information is very much dependent on what aspect of the quantum state of light one is interested in.
\end{abstract}

\maketitle

\section{Introduction}

Single photons play an essential role in both fundamental and applied physics. 
On the one hand, for example, it has been theoretically argued~\cite{Hardy1994}  and experimentally verified~\cite{Hessmo2004} that nonlocality may be an intrinsic characteristic of a single electromagnetic field excitation.
On the other, due to their weak interaction with the environment, single photons are good candidates for quantum key distribution where security may be monitored in real time~\cite{Kupko2019}.
The full range of applications beyond those mentioned here stoke the quest for a perfect single-photon source that emits indistinguishable single photons on demand at a high repetition rate that can be collected or coupled to optical channels with high efficiency~\cite{Reimer2019}.
In a more realistic approach, different applications require different characteristics~\cite{Lachman2019} but all call for a high-quality single-photon source.
The characterization of such sources is a relatively open problem. 
The current standard is based on the second-order correlation function,
$g^{(2)}(\tau)$~\cite{HBT1956,Glauber63,Sudarshan63} and assumes that single-photon states produce zero-delay values, $\tau = 0$, that are less than one-half, $g^{(2)}(0) <1/2$~\cite{Michler2000,Vuckovic2012}. 
In this notation, $g^{(2)}(0)$ is defined via creation(annihilation) operators $\hat a^\dagger(\hat a)$ of a single-mode field in the steady state as
\begin{equation}
g^{(2)}(0)=\frac{\langle\hat a^{\dagger2}\hat a^2\rangle}{\langle\hat a^\dagger\hat a\rangle^2}.
\end{equation}
As this manuscript deals only with $g^{(2)}(\tau=0)$, we will omit the argument from now on.

Recently, two independent results shed new light on what can be estimated from low values of $g^{(2)}$.
Zubizarreta et al.~\cite{Zubizarreta2017} showed that sub-Poissonian light, $g^{(2)}<1$, implies a limited average photon number $N=\langle\hat a^\dagger\hat a\rangle$ and derived a hard lower bound of $g^{(2)}$ for any given $N$.
In particular, a value $g^{(2)}<1/2$ implies a quantum state with $N<2$.
Gr\"unwald~\cite{g2paper} showed that $g^{(2)}<1/2$ is sufficient to verify a non-zero single-photon projection.
However, it can have an arbitrary amplitude.
In fact, $g^{(2)}$ does help discriminate multi-photon contributions from single- \textsl{and} zero-photon contributions within the quantum state.
It provides relative bounds for the ratio of single-to-multi-photon projection.
Additional information, therein given in the form of the vacuum component, is required to obtain absolute bounds.

These results yield a somewhat surprising picture. 
The standard criterion $g^{(2)}<1/2$ is enough to witness the existence of a single-photon projection within the quantum state of light. These quantum states form a subset of sub-Poissonian (i.e. nonclassical) states for which $g^{(2)}<1$. 
Yet, estimating the bounds of the single-photon projection requires additional information. 
In turn, knowledge of this additional parameter extends the range of applicable states. 
For example, for sources with 50\% or more vacuum projection some classical states can be addressed such as low-excitation coherent and thermal states~\cite{g2paper}. 
Hence, $g^{(2)}$ as a source of information about the single-photon content of a quantum state of light is not restricted to nonclassical states.

Our aim here is to analyze how access to both the zero-time delay second-order correlation $g^{(2)}$ and the mean photon number $N$ allows us to obtain knowledge about the single-photon projection (SPP) and the single-to-multi-photon projection ratio (SMPPR).
We derive a new set of bounds for these quantities combining the results in the literature~\cite{Zubizarreta2017,g2paper} and show that an effective second-order correlation function $\tilde g_N^{(2)}$ emerges, similar to the effective second-order correlation function $\tilde g^{(2)}_0$ defined in~\cite{g2paper}. 
This quantity allows us to characterize the SPP and SMPPR for a range of states beyond that of the standard criterion $g^{(2)}<1/2$; in particular, classical states. 
Comparing the two effective correlation functions we find that the SMPPR has tighter lower bounds using $\tilde g^{(2)}_N$.
In contrast, absolute bounds are described better by $\tilde g_0^{(2)}$ at least for low excitations states. 
In general, more states can be analyzed with our new criteria compared to those in the literature~\cite{g2paper}.

The article is organized as follows. In Sec.~\ref{sec.previous}, we will give a brief review of major results from the literature~\cite{Zubizarreta2017,g2paper}. 
Afterwards, we present our derivation of the bounds for SPP and SMPPR based on $g^{(2)}$ and $N$ in Sec.~\ref{sec.analysis}. 
In Sec.~\ref{sec.comp} we compare the quality of the different criteria derived here and in~\cite{g2paper}. 
We give some examples for applications in Sec.~\ref{sec.appl}. 
Finally, in Sec.~\ref{sec.concl}, we present some conclusions and an outlook.


\section{State of the art}\label{sec.previous}

Let us briefly review the details of two recent papers~\cite{Zubizarreta2017,g2paper} relevant to our work.
For a given average photon number $N$ larger than one, there is a nonzero lower bound on the value of the zero-time delay second-order correlation function~\cite{Zubizarreta2017},
\begin{equation}
g^{(2)} \geq \frac{\lfloor N \rfloor(2N-\lfloor N \rfloor-1)}{N^2},\label{eq.Zubi}
\end{equation}
as higher Fock states must be occupied. 
We define the floor function $\lfloor x \rfloor$ as the largest integer less than $x$. 
This hard boundary means that for any given $N$ there exists a quantum state with $g^{(2)}$ equal to the right-hand-side of Eq.~(\ref{eq.Zubi}). 
Despite the discontinuity of the floor function, the lower bound is continuous and monotonically increasing. 
In the limit $N\to\infty$, the lower bound reaches one, as, for example, coherent states with $g^{(2)}=1$ are independent of $N$. 
It is relevant to our discussion that the average photon number is limited by two $N<2$ for $g^{(2)}<1/2$.

In comparison,~\cite{g2paper} is explicitly concerned with the SPP in a general quantum state and its relation to low $g^{(2)}$.
We start by splitting the quantum state of light into three parts,
\begin{equation}
|\psi\rangle = \sqrt{p_0}|0\rangle + \sqrt{p_1}|1\rangle+\sqrt{q}|\psi_2\rangle, \label{eq:state}
\end{equation} 
where the projections onto the vacuum, single- and multi-photon states are given by $p_{0}$, $p_{1}$ and $q$ in that order. 
All expectation values discussed in~\cite{g2paper} and this work, $\{ g^{(2)},N,p_k \}$, with the latter being the Fock-state projection $p_k=|\langle k|\psi\rangle|^2$, are diagonal in the Fock basis. 
Thus, we can limit the discussion to states of the type given by Eq.~(\ref{eq:state}) without loss of generality.
When there is no vacuum projection, $p_0=0$, an increase of the multi-photon projection $q$ aligned with a decreasing SPP $p_1$ leads to an expected increase of the second-order correlation $g^{(2)}$. 
Less obvious, the vacuum contribution $p_{0}$ can also yield a higher value of $g^{(2)}$ while, in reality, lowering both the single- and multi-photon projection. 
Incoherently mixing vacuum into a quantum state without vacuum projection preserves the SMPPR $p_1/q$. 
This leads to an effective second-order correlation function
\begin{equation}
\tilde{g}_0^{(2)} = (1-p_0)g^{(2)}, \label{eq.g(2)0}
\end{equation}
that accounts for the effect of the vacuum projection $p_0$.
It is possible to derive a lower bound on the SMPPR that depends solely on $\tilde g_0^{(2)}$,
\begin{equation}
\frac{p_1}{q} \geq \frac{2\sqrt{1-2\tilde{g}_0^{(2)}}}{1-\sqrt{1-2\tilde{g}_0^{(2)}}}.\label{eq:C}
\end{equation}
The right-hand side is monotone with respect to the effective second-order correlation function $\tilde g^{(2)}_0$.
Hence, even without knowledge of the vacuum projection this ratio has a lower bound based solely on $g^{(2)}<1/2$. 
Furthermore, when this ratio is expected to be at least some value $\theta$, the effective second-order correlation function is bounded from above by
\begin{equation}
\tilde{g}^{(2)}_0 \leq \frac{2(\theta+1)}{(\theta+2)^2}\label{eq.theta_0}.
\end{equation}
In contrast to these relative bounds, the absolute bounds on the SPP $p_1$ require individual knowledge of the vacuum projection $p_0$ and the second-order correlation function $g^{(2)}$,
\begin{equation}
(1-p_0) \geq p_1 \geq (1-p_0)\frac{2\sqrt{1-2\tilde{g}_0^{(2)}}}{1+\sqrt{1-2\tilde{g}_0^{(2)}}}.\label{eq.abs_bounds_0}
\end{equation}


\section{Single-photon projection analysis}\label{sec.analysis}

The average photon number $N$ and the second-order correlation function $g^{(2)}$ have encoded information on all the Fock-state projections $p_k$ of a quantum state. 
Together with the completeness relation $\sum_{k=0}^\infty p_k=1$, they can be used to obtain well-defined lower boundaries on the single-photon projection $p_1$. 
Akin to~\cite{g2paper}, we define the contributions to $N$ and $g^{(2)}$ from the multi-photon projection $|\psi_2\rangle$, which have fixed lower bounds
\begin{eqnarray}
n_2=&\langle\psi_2|\hat a^\dagger\hat a|\psi_2\rangle &\geq 2, \\
g_2=&\dfrac{\langle\psi_2|\hat a^{\dagger2}\hat  a^2|\psi_2\rangle}{n_2^2} &\geq \dfrac{1}{2}, \label{eq.n2g2bound}
\end{eqnarray}
as $|\psi_2\rangle$ contains no projection on vacuum and the single-photon Fock state.
Both lower bounds are attained for $|\psi_2\rangle=|2\rangle$.
For the general state $|\psi\rangle$ we can write,  
\begin{eqnarray}
N =& \langle\psi|\hat a^\dagger\hat a|\psi\rangle &= p_1 + n_2q,\label{eq:N}\\
g^{(2)} =& \dfrac{\langle\psi|\hat{a}^{\dagger 2}\hat{a}^2|\psi\rangle}{\langle\psi|\hat a^\dagger\hat a|\psi\rangle^2} &= \dfrac{qn_2^2g_2}{N^2}.\label{eq:gt}
\end{eqnarray}
In order to recover information on $p_1$, we use Eq. \eqref{eq:N} to substitute $qn_2$ on Eq. \eqref{eq:gt} and solve for the SPP, 
\begin{equation}
p_1 = N \left( 1 - \frac{Ng^{(2)}}{n_2g_2} \right).\label{eq:p1}
\end{equation}
This is an exact relation that connects the SPP to $N$ and $g^{(2)}$. 
It states that for null SPP the product $Ng^{(2)}$ of the overall quantum state must be equal to its multi-photon counterpart $n_2 g_2$. 
If the state has neither vacuum nor SPP, this is obvious. 
If $p_0>0$, then the overall state represents vacuum mixing to the state $|\psi_2\rangle$, as described before Eq.~(\ref{eq.g(2)0}), and $N \to n_2(1-p_0)$ and $g^{(2)}\to g_2/(1-p_0)$ cancel the vacuum contribution in the product.

The average photon number $N$ and second-order correlation function $g^{(2)}$ are measurable quantities, but $n_2$ and $g_2$ are difficult to obtain. 
Using their lower bounds from Eq.~(\ref{eq.n2g2bound}) and combining them with Eq.~(\ref{eq:N}), we can derive absolute bounds,
\begin{equation}
N \geq p_1 \geq N(1-Ng^{(2)}),\label{eq.abs_bounds_N}
\end{equation}
just in terms of $N$ and $g^{(2)}$. 
The lower bound is identical to the exact value for $p_1$ in the case of no more than two-photon projection as derived in~\cite{Zubizarreta2017}. 
This is a direct consequence of the quasiconcaveness of both $N$ and $g^{(2)}$ as a function of the quantum state $\hat\varrho$~\cite{g2paper,gk2019}. 
Considering that potential single-photon emitters usually are limited to low Fock-state projections, it is worth mentioning that our lower bound is exact within this restriction and only higher Fock states $|n\geq 3\rangle$ yield deviations.

The lower bound on the SMPPR, 
\begin{equation}
\frac{p_1}{q} = \frac{n_2^2g_2}{N g^{(2)}}-n_2 \geq 2\left(\frac{1}{Ng^{(2)}}-1\right).\label{eq.rel_bounds_N}
\end{equation}
follows from Eq.~\eqref{eq:p1} by substituting the $N$ in front of the brackets according to Eq.~\eqref{eq:N}.
In this case, we only have one factor of $Ng^{(2)}$ that scales the lower bound of the ratio. 
This factor is reminiscent of the effective second-order correlation function $\tilde g_0^{(2)}$ as derived in~\cite{g2paper}. 
In particular, when analyzing previous results therein, for $N<1$ we could use $p_0\geq 1-N$ and thus used $Ng^{(2)}$ as upper bound on $\tilde g_0^{(2)}$. 
The only difference here is that we are not limited to $N<1$ but only to $Ng^{(2)}<1$. 
It thus seems appropriate to propose another effective second-order correlation function,
\begin{equation}
\tilde{g}_N^{(2)} = Ng^{(2)}.\label{eq.g2N}
\end{equation}

A few comments are in order. 
Both the absolute and relative lower bounds become relevant i.e. real and positive, for $\tilde g^{(2)}_N<1$, in difference to $\tilde g_0^{(2)}$, which must have a value below $1/2$.
However, using the major result from~\cite{Zubizarreta2017}, that for states with $g^{(2)}<1/2$ it follows that $N<2$, we directly recover that a nonzero SPP is generally given for $g^{(2)}<1/2$. The absolute lower bound on $p_1$, on the other hand, is scaled by $N$ and can be arbitrarily small for low excitation $N \ll 1$.

Both bounds are also hard boundaries.
If no more than two-photon projections are given, they are exact equalities as $p_1$ and $p_2=q$ are fully described by $N$ and $g^{(2)}$ together with the completeness relation for the density matrix~\cite{Zubizarreta2017}. 
Likewise, when we increase the contribution from higher Fock states, the value of $q$ for a given $g^{(2)}$ and $N$ decreases. 
Hence, there is no upper bound on the SMPPR, and the absolute upper bound for $p_1$ is only exact for $q=0$.

For a vanishing effective second-order correlation, $\tilde g^{(2)}_N\to0$, we obtain the correct statement that $N=p_1$, while the multi-photon contribution vanishes and $p_1/q$ diverges. 
In general, Eq.~(\ref{eq.rel_bounds_N}) monotonically increases with decreasing $\tilde g^{(2)}_N$. 
Hence, we can determine an upper bound on the effective second-order correlation function to guarantee a desired SMPPR $\theta$,
\begin{equation}
\tilde{g}_N^{(2)} \leq \frac{1}{\tfrac{1}{2}\theta+1}=\frac{2(\theta+2)}{(\theta+2)^2}.\label{eq.theta_N}
\end{equation}
This bound is looser than that for vacuum as additional information, Eq.~(\ref{eq.theta_0}), indicating that using a photon-number based effective correlation function may be advantageous when it comes to the SMPPR.

Due to the completeness relation, the SMPPR implies an upper bound on the multi-photon projection,
	\begin{equation}
	q\leq\frac{1}{\tfrac{2}{\tilde g^{(2)}_N}-1}=\frac{\tilde g^{(2)}_N}{2-\tilde g^{(2)}_N},
	\end{equation}
where we used $p_1\leq 1-q$.
Correspondingly, a lower bound on the sum of vacuum and SPP follows,
	\begin{equation}
	p_0+p_1=1-q\geq \frac{2(1-\tilde g^{(2)}_N)}{2-\tilde g^{(2)}_N}.
	\end{equation}
This reveals, in accordance with the statement from the introduction, that $g^{(2)}$ does not actually seperate single- from multi-photon projection, but only both single- and zero-photon projection from multi-photon projections. To avoid confusion, $p_1>0$ is still needed to allow $g^{(2)}$ to fall below 1/2. However, using the effect of vacuum, one can lower $p_1$ for any given $g^{(2)}$ arbitarly close to zero, cf.~\cite{g2paper}. Hence, the Fock-state projection for which $g^{(2)}<1/2$ without further information implies a nonzero lower bound is $p_0+p_1$.

One fundamental difference compared to $\tilde g^{(2)}_0$ is that ignorance of the additional parameter $N$ can not be as easily compensated as for the vacuum projection $p_0$. 
In the latter case, the monotonicity of the relative lower bound allowed us to set $p_0=0$ and hence $\tilde g^{(2)}_0=g^{(2)}$ and still obtain a lower bound for $p_1/q$, albeit not as good as for $\tilde g^{(2)}_0$. 
In the case presented here, one would have to set $N=1$ for that limit but this would break the lower boundary aspect. 
Consider a state with no Fock-state projections above $n=2$.
As shown above, Eq.~(\ref{eq.rel_bounds_N}) becomes an identity. 
Setting $\tilde g^{(2)}_N=g^{(2)}$ would yield the correct identity for $N=1$, a lower bound still for $N>1$, but an upper bound on the single-to-multi-photon projection ratio for $N<1$. 

Instead, one can again use the result from~\cite{Zubizarreta2017} and set $N=2$, its maximum value for $g^{(2)}\leq1/2$. This leads to obtain
	\begin{align}
	\frac{p_1}{q}\geq&2\left(\frac{1}{2g^{(2)}}-1\right)=\frac{1}{g^{(2)}}-2,\\
	p_0+p_1\geq&\frac{1-2g^{(2)}}{1-g^{(2)}},\\
	N\geq p_1\geq&N(1-2g^{(2)}).
	\end{align}
However, in these cases, we do not preserve a subset of states, where the formulas would still be exact. For the vacuum-based cases, we would recover exact equalities if the state were limited to projections on single- and two-photon components.
Here, the only state yielding an equality is the Fock state $|2\rangle$ that implies $p_1=0$.

To get some deeper insight into the physical meaning of $\tilde g^{(2)}_N$, it is convenient to express this quantity via $N$ and either the photon number variance $(\Delta N)^2$ or the Mandel-$Q$ parameter~\cite{Mandel79},
\begin{equation}
\tilde g^{(2)}_N=N-1+\frac{(\Delta N)^2}{N}=N+Q.
\end{equation}
It is clear that $N-1$, given for Fock states, is a lower bound for $\tilde g^{(2)}_N$. 
Hence, there is no state with an average photon number $N\geq2$ that can be identified with our criteria. 
This also follows from the fact that $N\geq2$ requires $g^{(2)}\geq1/2$ and thus $\tilde g^{(2)}_N\geq1$. 
However, there is nothing that hinders us from assuming $N<2$ and $g^{(2)}>1/2$, as long as $\tilde g^{(2)}_N<1$. 
Thus, we can again extend the range beyond states usually addressed by $g^{(2)}<1/2$ and consider even some classical states. 

In short, we can distinguish three sets of states: $M_1$, the set of states $\hat\varrho$ for which $g^{(2)}<1/2$; $M_2$, the set of states for which $\tilde g^{(2)}_N<1$; and $M_3$, the set of states for which $N<2$. 
The first two sets constitute states for which our criteria detect nonzero SPP, the third one does not. 
The relation between these sets is given by
\begin{equation}
M_1\subset M_2\subset M_3.
\end{equation}
A simple example for states in the difference of sets $M_2-M_1$ and $M_3-M_2$ are coherent states with $N<1$ and $1\leq N<2$, respectively. The details of these kinds of states will be discussed in the next section.


\section{Comparison between effective correlation functions}\label{sec.comp}

Having established a connection between the second-order correlation function $g^{(2)}$, the average photon number $N$, and the single-photon projection $p_1$ akin to the results of~\cite{g2paper}, an apparent question to ask is, which of the two descriptions yields better boundaries when analyzing an unknown quantum state. 
More precisely, given a light source that is potentially relevant as a single-photon source, which approach yields tighter bounds for either the absolute SPP, or the SMPPR. 

\begin{figure}[ht]
	\includegraphics[width=8cm]{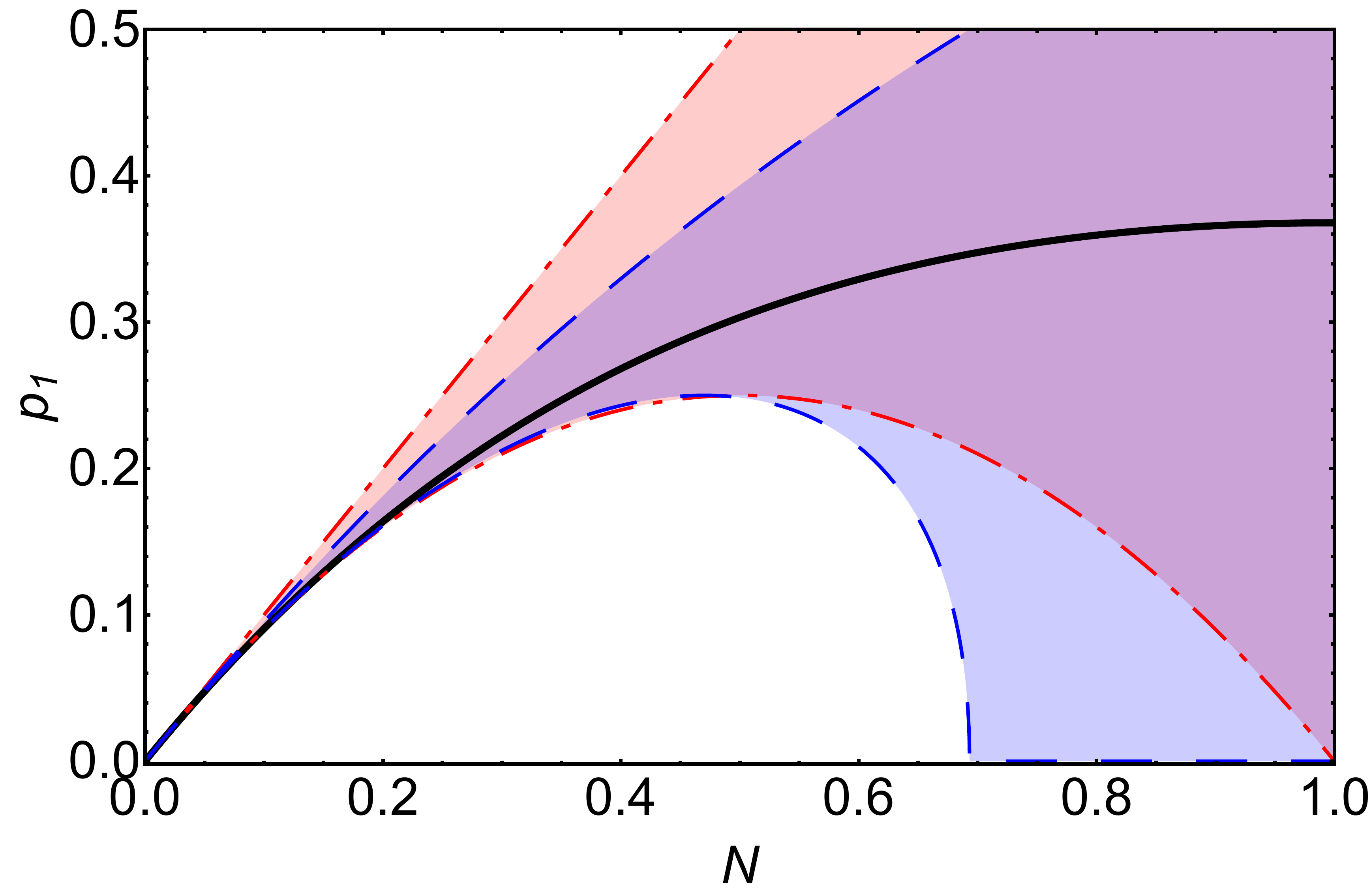}
	\hfill
	\includegraphics[width=8cm]{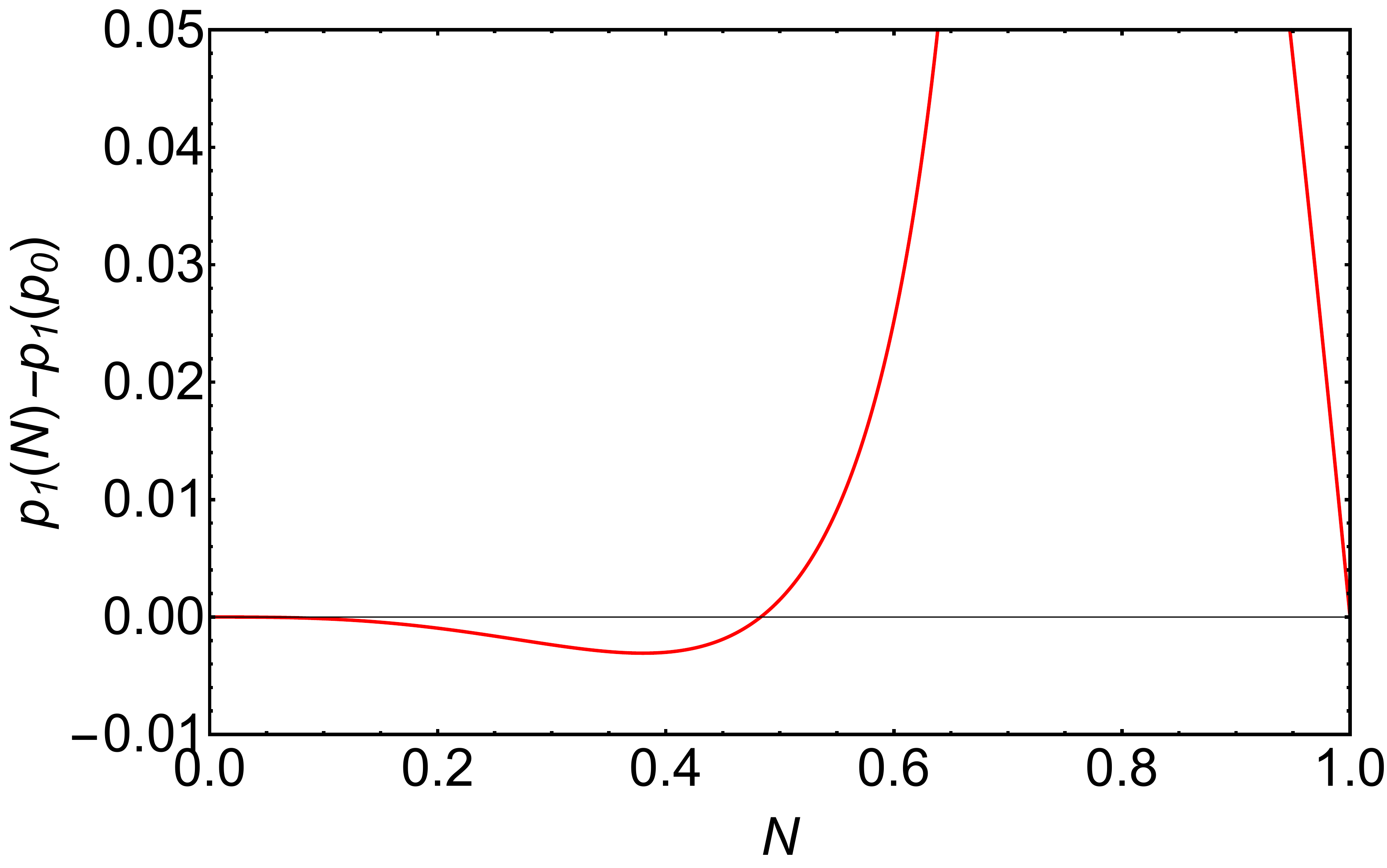}
	\caption{(Color online) Top: Bounds on single-photon projection $p_1$ of coherent states over $N$ (exact value shown as black thick solid line) as given by Eq.~(\ref{eq.abs_bounds_0}) (blue dashed lines) and Eq.~(\ref{eq.abs_bounds_N}) (red dot-dashed line). Bottom: Difference between lower bound $p_1(N)$ based on $N$ and $p_1(p_0)$ based on $p_0$.}\label{fig.CohAbs0N}
\end{figure}

\begin{figure}[ht]
	\includegraphics[width=8cm]{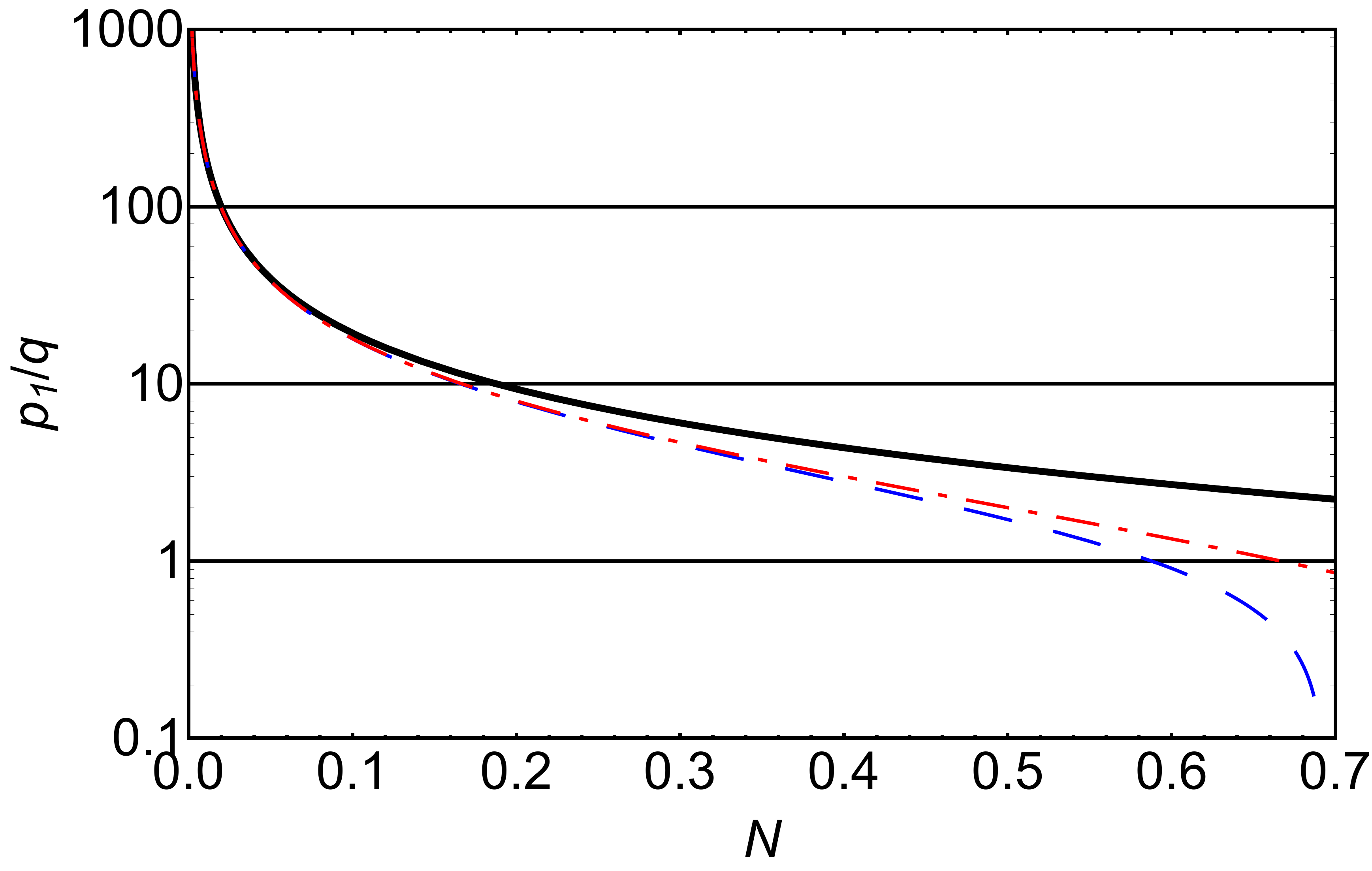}
	\hfill
	\includegraphics[width=8cm]{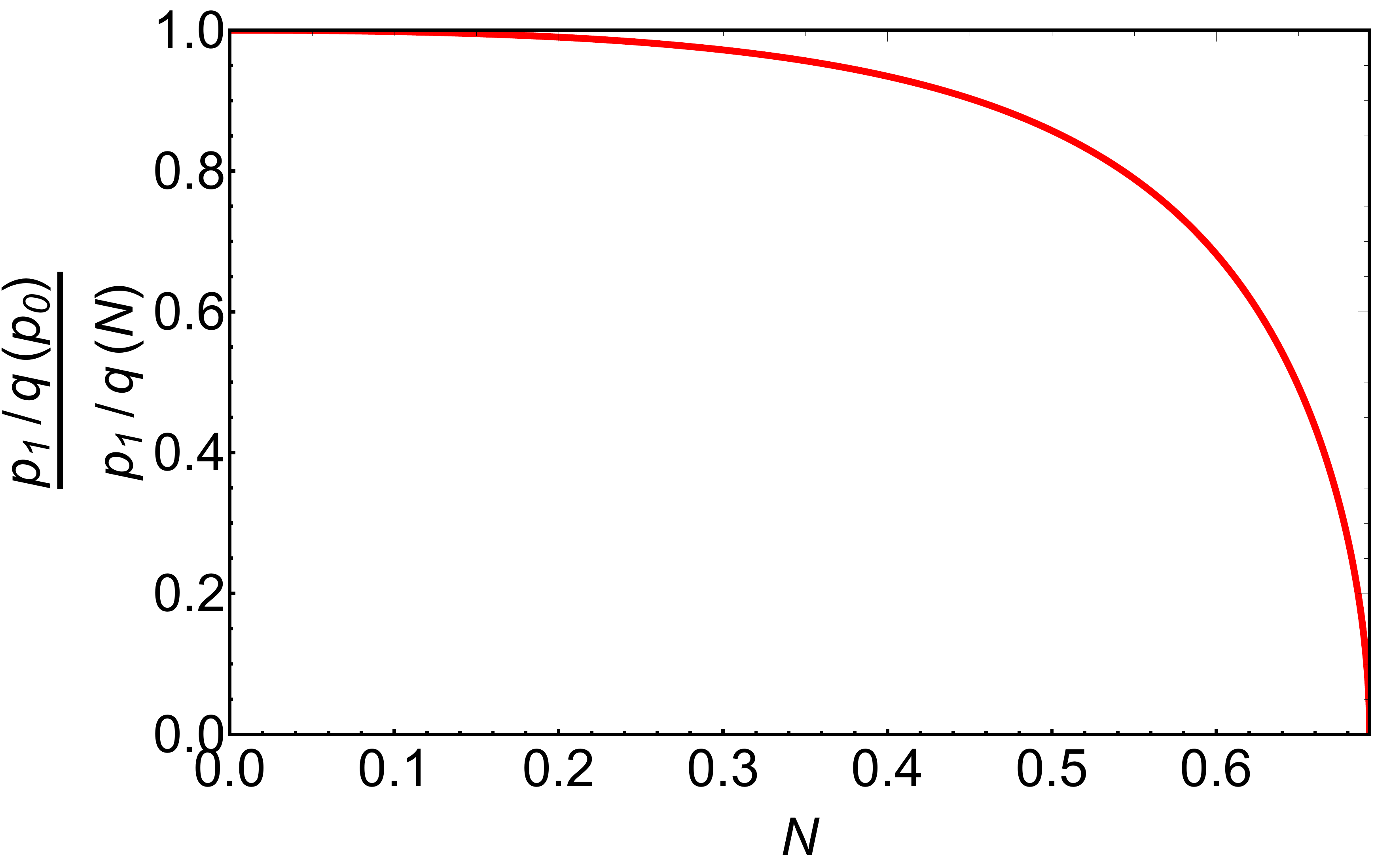}
	\caption{(Color online) Top: Lower bounds on single-to-multi-photon projection ratio $p_1/q$ for coherent states over $N$. The lines are the same as for the absolute bounds on $p_1$. Bottom: Ratio of the lower bound $p_1/q(p_0)$ based on $p_0$ and $p_1/q(N)$ based on $N$.}\label{fig.CohRel0N}
\end{figure}

For this purpose, let us first analyze the well-known coherent  $(g^{(2)}=1)$, and thermal $(g^{(2)}=2)$ quantum states~\cite{WelVo}. 
The introduction of the effective second-order correlation functions $\tilde g^{(2)}_{0,N}$ extends the range of states, which can be identified as having a nonzero SPP, including low-excitation classical states. 
For vacuum projection as additional information, the range of validity, $\tilde g^{(2)}_0<1/2$ yields bounds of $N<\ln(2)$ and $N<1/3$ for coherent and thermal states, respectively. 
In comparison, for the average photon number itself, confer  Eq.~(\ref{eq.abs_bounds_N}), these bounds are shifted up to $N<1$ and $N<1/2$, respectively. 
Thus, using the average photon number instead of the vacuum projection allows to identify a larger subset of coherent and thermal states that contain a nonzero SPP.

However, the question of range of validity is independent of the actual bounds imposed on the absolute and relative projections. 
Thus, in Fig.~\ref{fig.CohAbs0N}, we depict the upper and lower bounds for $p_1$ according to Eqs.~(\ref{eq.abs_bounds_0}) and~(\ref{eq.abs_bounds_N}) for coherent states. 
One can see that, for sufficiently low excitation, both the lower and upper bounds are tighter using $p_0$ instead of $N$ as additional parameter. 
Hence, despite $\tilde g^{(2)}_N$ identifying a larger set of states, the absolute bounds on the SPP are not as good as with the vacuum projection.

\begin{figure}[ht]
	\includegraphics[width=8cm]{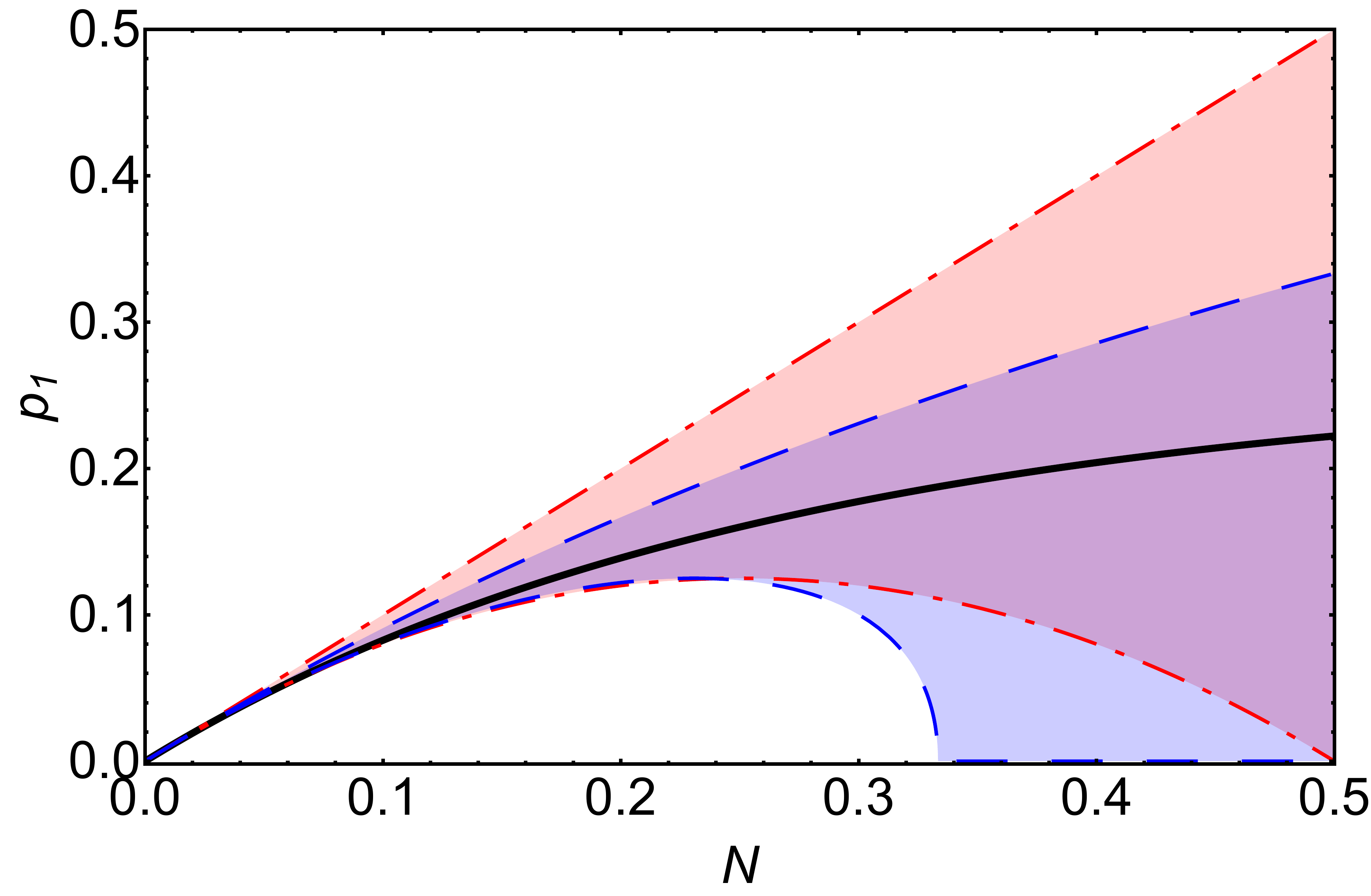}
	\hfill
	\includegraphics[width=8cm]{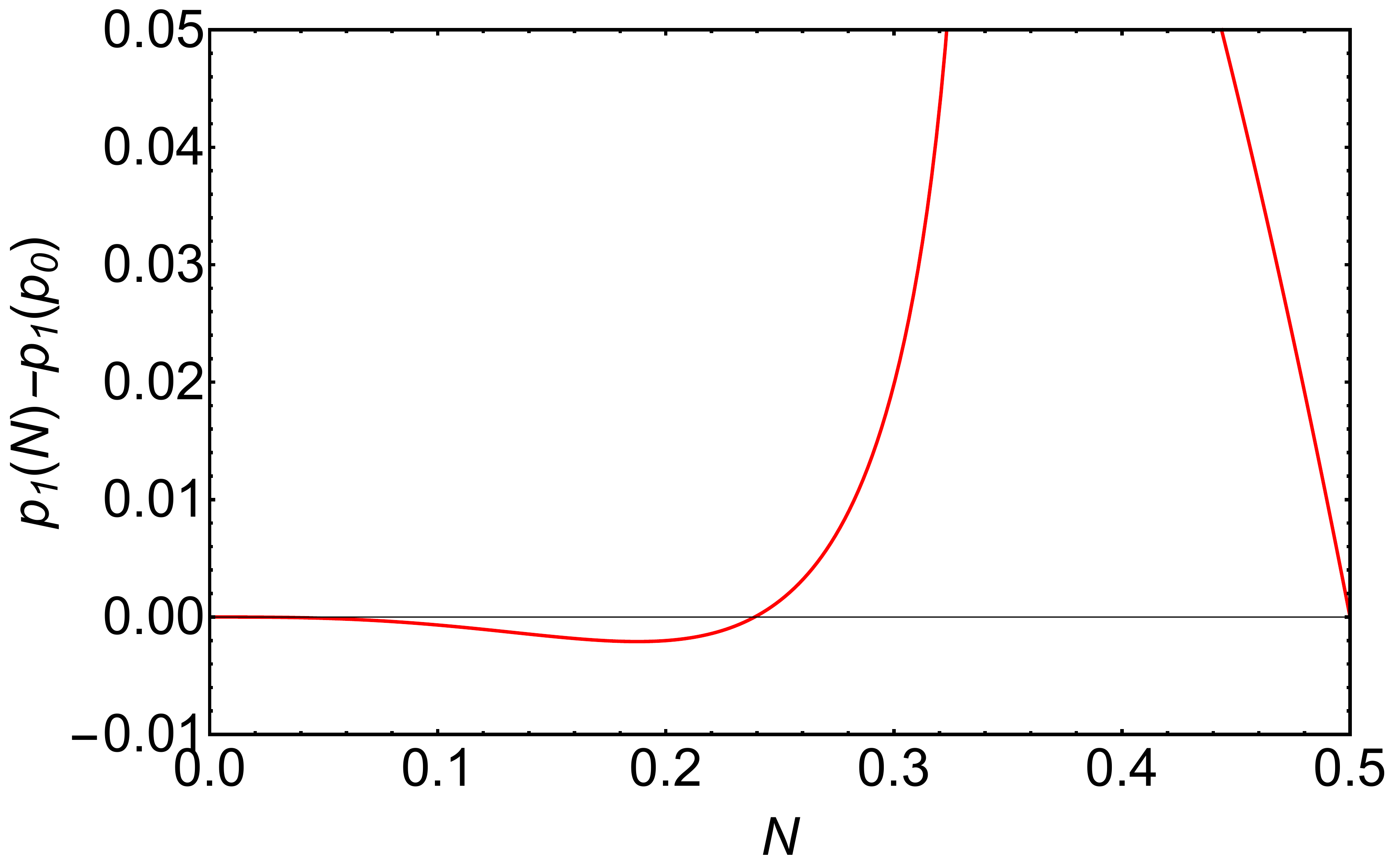}
	\caption{Same as in Fig.~\ref{fig.CohAbs0N}, but for thermal states.}\label{fig.ThermAbs0N}
\end{figure}

\begin{figure}[ht]
	\includegraphics[width=8cm]{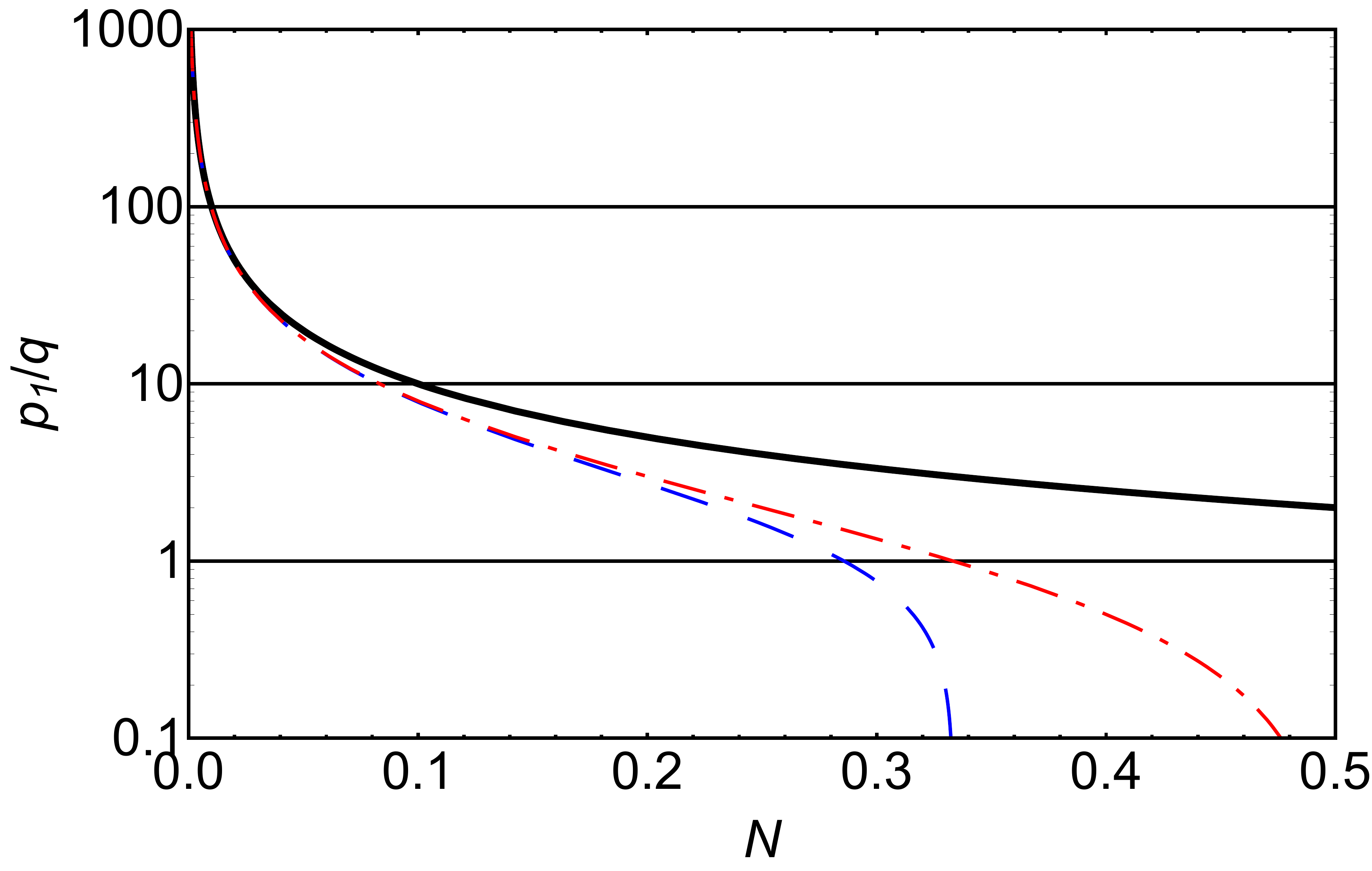}
	\hfill
	\includegraphics[width=8cm]{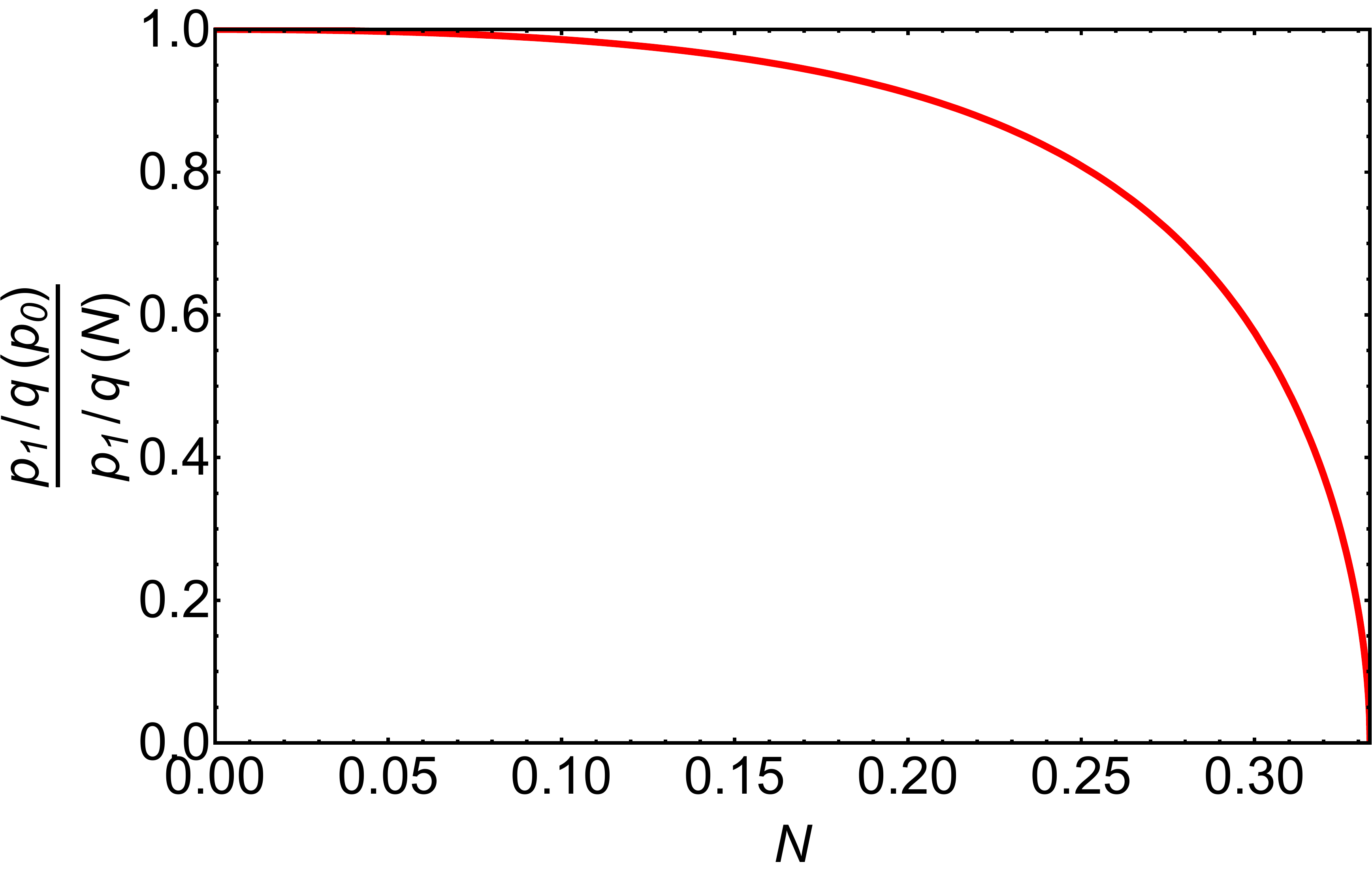}
	\caption{Same as in Fig.~\ref{fig.CohRel0N}, but for thermal states.}\label{fig.ThermRel0N}
\end{figure}

The picture changes drastically when one considers the relative lower bounds as depicted in Fig.~\ref{fig.CohRel0N}. 
Not only is the range of coherent states for which a nonzero lower bound can be concluded larger.
The bound itself is tighter for any excitation. 
Hence, we have the inverted scenario for relative bounds. 
Moreover, all of these qualitative comparisons also apply for thermal states, see Figs.~\ref{fig.ThermAbs0N},\ref{fig.ThermRel0N}.

This similar behaviour of bounds for coherent and thermal states motivates a question on whether one can show which additional information, the vacuum projection or the average photon number, yields preferable results. 
It turns out this is indeed possible to some extent.
Starting from the absolute upper bound, one can easily show that
\begin{equation}
1-p_0=p_1+q\leq p_1+n_2q=N.
\end{equation}
The absolute upper bound on $p_1$ will always be better described by the vacuum-induced bound than by the average photon number. 
A similar result can be shown for the lower bounds. 
For large vacuum and low average photon number, we have $1-p_0\approx N$ and $g^{(2)}\ll 1/2$. 
Developing the right-hand side of Eq.~(\ref{eq.abs_bounds_0}) for low $g^{(2)}$, we find
\begin{equation}
p_1\geq(1-p_0)\left(1-\tfrac{(1-p_0)g^{(2)}}{2}\right).
\end{equation}
In comparison, using the above approximation for $N \ll 1$ on the right-hand side of Eq.~(\ref{eq.abs_bounds_N}) yields
\begin{equation}
p_1\geq(1-p_0)\left[1-(1-p_0)g^{(2)}\right].
\end{equation}
When vacuum has a dominant contribution and $N$ is significantly below unity, one can expect that using criteria from~\cite{g2paper} should be more advantageous than using those derived here. 
However, when there is little to no vacuum projection, the criteria including average photon number may become better. 

For the relative lower bounds, we use an expansion of the right-hand-side of Eq.~(\ref{eq:C})~\cite{g2paper},
\begin{equation}
\frac{2\sqrt{1-2\tilde g^{(2)}_0}}{1-\sqrt{1-2\tilde g^{(2)}_0}}\leq \frac{2}{\tilde g^{(2)}_0}-3-\frac{\tilde g^{(2)}_0}{2}.
\end{equation}
As we are only interested in the sufficiently low excited cases, where $\tilde g^{(2)}_N\approx\tilde g^{(2)}_0$, we can easily compare this upper bound with the exact formula for the average-photon-number based criterion, Eq.~(\ref{eq.rel_bounds_N}). 
It is obvious in this notation that the average-photon number will always yield a better lower bound on the SMPPR than the vacuum projection $p_0$.

Finally, we discuss the range of states where a nonzero SPP can be detected. 
The results from Figs.~\ref{fig.CohAbs0N},\ref{fig.ThermAbs0N} indicate that the set of states covered by our criterion $\tilde g^{(2)}_N<1$ is always larger than those described by $\tilde g^{(2)}_0<1/2$. 
We prove this conjecture by means of an indirect proof, showing that a quantum state, where $\tilde g^{(2)}_N\geq1$ but  $\tilde g^{(2)}_0<1/2$ would require negative multi-photon projections $p_n<0$ with $n\geq2$. 
	
For the sake of notation, we denote for the rest of this section $\tilde g^{(2)}_N=g\geq1$.
We can write the explicit formula for our effective second-order correlation,
	\begin{align}
	\tilde g^{(2)}_N=&\frac{\sum\limits_{n=2}^\infty n(n-1)p_n}{\sum\limits_{n=1}^\infty np_n}=g,
	\end{align}
and calculate the SPP,
	\begin{align}
	p_1=&\frac{1}{g}\sum\limits_{n=2}^\infty n(n-1-g)p_n>0.\label{eq.p1g}
	\end{align}
In order to detect a nonzero SPP, a ratio of the two effective correlation functions follows as
	\begin{align}
	\frac{\tilde g^{(2)}_N}{\tilde g^{(2)}_0}=&\frac{N}{1-p_0}>2g.	
	\end{align}
Solving again for $p_1$ and inserting Eq.~(\ref{eq.p1g}), we obtain a condition for the multi-photon projections,
	\begin{equation}
	\sum\limits_{n=2}^\infty(n-2g)p_n>(2g-1)p_1=\left(2-\frac{1}{g}\right)\sum\limits_{n=2}^\infty n(n-1-g)p_n.\label{eq.sum_g}
	\end{equation}
Bringing the sums to the right side and simplifying finally gives
	\begin{equation}
	0>\sum\limits_{n=2}^\infty(n-1)\left[n\left(2-\frac{1}{g}\right)-\left(1+\frac{1}{g}\right)\right]p_n.\label{eq.sum_g2}
	\end{equation}
As the $p_n$ are probabilities they cannot be negative, and at least one of the prefactors has to be below zero. 
For the lowest value of $g=1$, we get $n-2$ as the smaller factor, which is still greater or equal to zero due to the sum starting at $n=2$. 
For increasing $g$ this prefactor increases as well, thus yielding no negativities and, hence, Eq.~(\ref{eq.sum_g2}) is impossible to realize.

\section{Application}\label{sec.appl}

The fluorescent light emitted from a quantum dot is considered one of the best sources for single photons nowadays~\cite{Loredo16,Senellart17}.
The quantum state of these light sources is often assumed to be composed of a perfect single photon Fock state with a small coherent background from the exciting laser~\cite{Pelton02},
\begin{equation}
\hat\varrho=\tilde p_1|1\rangle\langle1|+(1-\tilde p_1)|\alpha\rangle\langle\alpha|.\label{eq.stateQD}
\end{equation}
Its full SPP is given by
\begin{equation}
p_1=\langle1|\hat\varrho|1\rangle=\tilde p_1+(1-\tilde p_1)N_\alpha\exp(-N_\alpha)\geq\tilde p_1.
\end{equation}
Depending on the amplitude of $\tilde p_1$ and the coherent-state average photon number $N_\alpha=|\alpha|^2$, vacuum contributions $\langle0|\hat\varrho|0\rangle=(1-p_1)\exp(-N_\alpha)$ may be negligibly small and, hence, the range of applicability of Eqs.~(\ref{eq:C}, \ref{eq.abs_bounds_0}) is limited to the same range as just using $g^{(2)}<1/2$. 
In that case, the standard limit to show nonzero SPP leads to a maximum background $N_\alpha$ as function of $\tilde p_1$ in the form
\begin{equation}
N_\alpha<\frac{\tilde p_1}{1+\tilde p_1}\left(1+\sqrt{\frac{2}{1-\tilde p_1}}\right).\label{eq.g(2)background}
\end{equation}
If the single-photon contribution from the quantum dot is very small, $\tilde p_1\ll1$, there is virtually no background allowed before it becomes undetectable with the standard criterion. 
On the other hand, it diverges in the limit $\tilde p_1\to1$ as there is no background left.

In contrast, our new criterion $ \tilde{g}_{N}^{(2)}= Ng^{(2)}<1$ leads to the upper bound,
\begin{equation}
N_\alpha<\frac{1}{2}\left(1+\sqrt{\frac{1+3\tilde p_1}{1-\tilde p_1}}\right).
\end{equation}
Even for no quantum-dot SPP, $\tilde p_1=0$, the background can be analyzed up to $N_\alpha=1$, as our method can detect the SPP of coherent states.
For Eq.~(\ref{eq.g(2)background}), this would require at least $\tilde p_1=50\%$ from the quantum dot. 
To visualize our advance with this additional knowledge, consider the following scenario. 
For any given $\tilde p_1$, we assume the background to be at the boundary given by the right-hand side of Eq.~(\ref{eq.g(2)background}). 
Thus, $g^{(2)}$ is always $1/2$ and no SPP can be detected. 
In contrast, due to the shift of the boundary caused by our additional information, we can give a nonzero lower bound on $p_1$ and compare with the exact value. 

\begin{figure}[ht]
	\includegraphics[width=8cm]{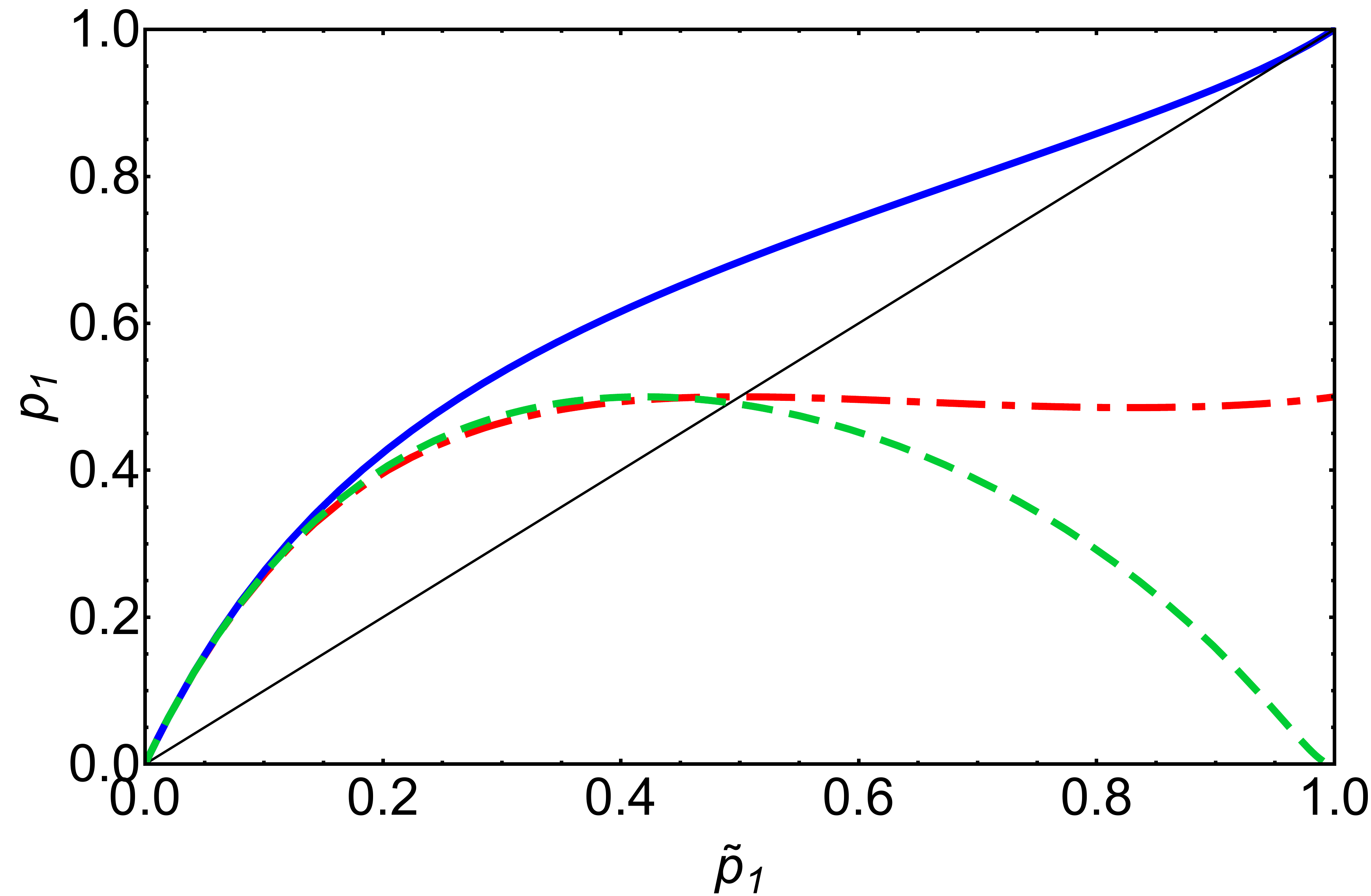}
	\caption{SPP of a designed state Eq.~(\ref{eq.stateQD}) where the coherent amplitude is adjusted so that $g^{(2)}=1/2$ for all $\tilde p_1$. The top solid blue curve shows the actual single-photon projection of this state, while the solid black diagonal is just the intended SPP from the quantum dot, $\tilde p_1$. The lower dot-dashed red curve depicts the lower bound of $p_1$ using the photon-number based criterion Eq.~(\ref{eq.abs_bounds_N}), while the dashed green curve shows the lower bound on $p_1$ using the vacuum-based criterion Eq.~(\ref{eq.abs_bounds_0}).}\label{fig.g2bg}
\end{figure}

The result of this setup is depicted in Fig.~\ref{fig.g2bg}. 
We see that the overall SPP is not only increasing, but strongly supplemented by the background, despite the second-order correlation function being 1/2. 
Moreover, our criteria not only give a lower bound that is very precise for strong background $\tilde p_1\ll1$. 
Also for large SPP, roughly half the projection is given as a lower bound. 
The missing contribution may be attributed to the large multi-photon projections in $|\alpha\rangle$. 
For comparison, we also computed the lower boundary based on the vacuum-criteria, Eq.~(\ref{eq.abs_bounds_0}). For $\tilde p_1\geq50\%$, the lower bound based on this criterion falls off, as vacuum becomes small, and hence we can not obtain more information than from just $g^{(2)}$.

\section{Conclusions and Outlook}\label{sec.concl}

We studied the information that can be gained about the SPP of an unknown quantum state of light from measuring its second-order correlation function $g^{(2)}$ and its average photon number $N$. 
If the effective second-order correlation function $Ng^{(2)}$ falls below unity, there is a nonzero but otherwise arbitrary SPP. 
We could derive a lower bound on the SMPPR based solely on the effective correlation function, as well as absolute lower and upper bounds on the SPP when using $g^{(2)}$ and $N$ individually. 
We compared these bounds to those previously derived using the vacuum projection instead of the average photon number as additional parameter.
The relative bounds are tighter using the average photon number, while the upper and in case of low excitation the lower bound on the SPP are better described by the vacuum contribution. 
The set of states detectable with our new criteria encompasses more states than those that can be analyzed with the vacuum based criteria. 
Our results show that depending on the actual goal when estimating the quality of a single-photon source, it may be different additional parameters that are of relevance.
	
It is noteworthy that all the criteria based on low $g^{(2)}$ derived here and in the literature are limited to states with comparably low excitation. 
Highly excited states, even with strong single-photon projections and sub-Poissonian photon distribution, can not be explained as the larger Fock-state contributions obscure the lower projections.
It seems questionable if a $g^{(2)}$-based approach can ever overcome this limitation.
	
Another open issue is the application of pulsed laser light, which is often used to overcome short coherence times of the emitter~\cite{Reindl2019}. 
In such a scenario the area of the pulse at zero delay is compared with other pulses to estimate $g^{(2)}$~\cite{Cosacchi2019}. 
However, the effect of these pulsed excitations is difficult to analyze, in particular with respect to its quantum content. 
Only recently was the description of quantum light in pulses put on solid theoretical footing~\cite{Kiilerich2019}. 
Hence, interpreting the quantum content of the pulses under study with respect to their single-photon projection will be a goal for future research.

\section*{Acknowledgments}
PG acknowledges financial support by the EU through the H2020-FETOPEN grant No. 800942 640378 (ErBeStA), and by the
DNRF through the Thomas Pohl Professorship maQma.


%

\end{document}